# Modulation of the ferromagnetic insulating phase in $Pr_{0.8}Ca_{0.2}MnO_3$ by Co substitution

T. Harada[1], I. Ohkubo[1, a)], M. Lippmaa[2], Y. Matsumoto[3], M. Sumiya[4], H. Koinuma[5], and M. Oshima[1, 6,]

[1]*Department of Applied Chemistry, The University of Tokyo, 7-3-1, Hongo, Bunkyo-ku, Tokyo, 113-8656, Japan*
[2]*Institute for Solid State Physics, The University of Tokyo, 5-1-5, Kashiwanoha, Kashiwa, Chiba, 277-8581, Japan*
[3]*Materials and Structures Laboratory, Tokyo Institute of Technology, 4259, Nagatsuta, Yokohama, 226-8503, Japan*
[4]*National Institute for Materials Science, 1-1 Namiki, Tsukuba, Ibaraki, 305-0044, Japan*
[5]*Graduate School of Frontier Science, The University of Tokyo, 5-1-5, Kashiwanoha, Kashiwa, Chiba, 277-8568, Japan*
[6]*CREST, Japan Science and Technology Agency (JST), 5 Sanbancho, Chiyoda-ku, Tokyo 102-0075, Japan*
[7]*The-University-of-Tokyo Synchrotron Radiation Research Organization, 7-3-1, Bunkyo-ku, Tokyo, 113-8656, Japan*

a) Author to whom correspondence should be adressed.
Electronic mail: **ohkubo@sr.t.u-tokyo.ac.jp**

Ferromagnetic insulator $Pr_{0.8}Ca_{0.2}Mn_{1-y}Co_yO_3$ ($0 \leq y \leq 0.7$) thin films were epitaxially grown on $(LaAlO_3)_{0.3}$-$(SrAl_{0.5}Ta_{0.5}O_3)_{0.7}$ (100) substrates by pulsed laser deposition. To probe the ferromagnetic insulator state of hole-doped manganites, the Co content dependences of the structural, magnetic, and transport properties were studied. Variation of lattice constant by the substitution of Co ions is well reproduced considering that divalent and trivalent Co ions substitute for Mn ions at the perovskite B-sites. For $0 \leq y \leq 0.3$, the Curie temperature, saturation magnetization, and magnetoresistance increase with increasing Co content, retaining the insulating properties. Detailed analyses of transport and magnetic properties indicate the contribution of both double exchange and superexchange interactions to the appearance of the ferromagnetic insulating phase. [DOI:    ]

Mixed-valent perovskite manganites $R_{1-x}A_xMnO_3$ have been widely studied because of their anomalous properties, such as colossal magnetoresistance (CMR)[1]. In hole-doped manganites, ferromagnetism is usually accompanied by metallic conduction of $e_g$ electrons according to the double exchange model. However, a ferromagnetic insulator (FI) state is observed in some lightly hole-doped materials, such as $La_{1-x}Sr_xMnO_3$ ($0.10 < x < 0.17$), $La_{1-x}Ca_xMnO_3$ ($x < 0.15$), $Nd_{1-x}Sr_xMnO_3$ ($0.2 < x < 0.28$), and $Pr_{1-x}Ca_xMnO_3$ (PCMO) ($0.15 < x < 0.3$)[1]. It is important to determine the origin of the FI state which remains poorly understood, in order to design functional materials such as multiferroics in which ferroelectric distortions can affect the magnetic ordering[2].

The origin of the FI state in $Pr_{1-x}Ca_xMnO_3$ ($0.15<x<0.3$) has been studied both from theoretical and experimental points of view[3-6]. In addition to the conventional double exchange and superexchange mechanisms, many other models have been proposed, such as the presence of several types of charge and orbital ordering (CO/OO)[3, 4] or orbital ordering[5, 6]. Mn-site substitution in the CO/OO state of $Pr_{1-x}Ca_xMnO_3$ ($x = 0.5$) is known to result in a collapse of long-range CO/OO ordering[7-9]. If the FI state of $Pr_{1-x}Ca_xMnO_3$ ($0.15 < x < 0.3$) is related to some CO state, Mn-site substitution may affect the CO state, resulting in drastic change of magnetic and transport properties. In this paper, we have studied the effect of Co substitution at the Mn-site on structural, ferromagnetic, and insulating properties of $Pr_{1-x}Ca_xMnO_3$ ($x = 0.2$) to seek the origin of the ferromagnetic insulator state in hole-doped manganites.

Epitaxial $Pr_{0.8}Ca_{0.2}Mn_{1-y}Co_yO_3$ (PCMCO) ($0 \leq y \leq 0.7$) thin films were grown on atomically flat (100)-oriented $(LaAlO_3)_{0.3}$-$(SrAl_{0.5}Ta_{0.5}O_3)_{0.7}$ (LSAT) substrates by pulsed laser deposition at a substrate temperature of 700 °C. Pure oxygen was constantly fed into the growth chamber to maintain an oxygen pressure of 30 mTorr. The epitaxial orientation and crystalline quality of the films were evaluated by X-ray diffraction (XRD). In order to determine the in-plane lattice constant, reciprocal space mappings (RSMs) were measured by four-circle X-ray diffraction (4c-XRD). Magnetic and transport properties were characterized using the Quantum Design magnetic property measurement system (MPMS) and physical property measurement system (PPMS), respectively.

Figure 1(a) shows the XRD pattern of a PCMCO ($y = 0.3$) thin film. The XRD pattern without secondary phase peaks and the atomically flat surface in atomic force microscopy (AFM) image shown in the inset of Fig. 1(b) show no signature of segregation. The in-plane and out-of-plane lattice constants of PCMCO ($0 \leq y \leq 0.7$), determined from RSMs around the PCMCO (303) and (300) peaks, are shown in Fig. 1(b). The in-plane lattice constant was almost constant regardless of Co content and close to the lattice constant of the LSAT substrate (green line), suggesting that all PCMCO films were epitaxially strained. The out-of-plane lattice constants linearly decrease with increasing the Co content, indicating successful substitution of Co ions for Mn ions at the perovskite B-site.

The magnetic field dependences of magnetization (M-H curves) are shown in Fig. 2(a). The M-H curve of a $Pr_{0.8}Ca_{0.2}MnO_3$ film shows a small coercive field and residual magnetization. The coercive fields of the PCMCO films increase drastically with increasing Co content. The Curie temperature ($T_C$) increase with Co content in the $0 \leq y \leq 0.3$ range, as shown in Fig. 2(b), (c). For $y > 0.5$, $T_C$ could not be determined because a clear ferromagnetic transition was not observed. To estimate the valence state of the substituting Co ions, the relative pseudo-cubic lattice constant ($\Delta d(y) = d(y) - d(y = 0)$) is plotted in Fig. 2(d). We employ Shannon's ion radii of $Co^{2+}$ (divalent, high spin, 0.745 Å), $Co^{II}$ (divalent, low spin, 0.65 Å), $Co^{3+}$ (trivalent, high spin, 0.61 Å), $Co^{III}$ (trivalent, low spin, 0.545 Å), $Co^{4+}$ (tetravalent, high spin, 0.53 Å), $Mn^{3+}$ (trivalent, high spin, 0.645 Å), and $Mn^{4+}$ (0.530 Å)[10] to estimate $\Delta d(y)$. The valency of Co doped in $Pr_{0.5}Ca_{0.5}MnO_3$ are known to be divalent according to the X-ray absorption spectroscopy (XAS) study on Co-doped $Pr_{0.5}Ca_{0.5}MnO_3$ [11]. As a result, Co substitution would change valency of $Mn^{3+}$ to be $Mn^{4+}$[11, 12]. Here, we assume that Co ions would be divalent also in PCMCO. On this assumption, Co substitution for Mn would result in decrease of two $Mn^{3+}$ and increase of one $Mn^{4+}$ and one divalent Co in low Co content region ($y<0.4$). Suppose all the Mn would become $Mn^{4+}$ in $y=0.4$, further substitution of the Mn for Co would result in decrease of one $Mn^{4+}$ and one divalent Co and increase of two trivalent Co. As shown in Fig. 2(d), experimentally observed Co content dependence of $\Delta d$ are close to the estimated lines for $Co^{2+}$ & $Co^{3+}$ (orange thick line) as well as $0.8Co^{3+}+0.2Co^{4+}$ (green dotted line). Co valency would be divalent considering the XAS study on Co-doped $Pr_{0.5}Ca_{0.5}MnO_3$[11], which favors the former estimation line. These discussions suggest that Co ions might exist as $Co^{2+}$ in low Co content region ($y<0.4$), and mixture of $Co^{2+}$ and $Co^{3+}$ in high Co content region ($y>0.4$). Though the spin state of trivalent Co cannot be determined from this simplified discussion, $Co^{3+}$ might exist considering the small energy difference (~30 meV) between $Co^{3+}$ and $Co^{III}$ reported in $La_{1-x}Sr_xCoO_{3-\delta}$[13]. Figure 2(e) shows the saturation magnetization relative to the non-substituted PCMO. The lines shown in the figure were calculated by counting unpaired spins. The experimental data increase more rapidly than all the calculated lines for various valence state of Co including $Co^{2+}$ & $Co^{3+}$ suggested from the lattice constant variation, indicating that the enhancement of magnetization cannot be explained simply by an increase of unpaired spins at the B-site ions, and enhancement of ferromagnetic interactions between B-site ions should be also considered.

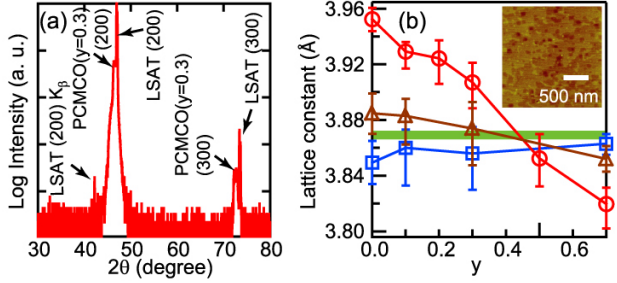

Fig. 1. Structural properties of epitaxial PCMCO thin films. (a) An XRD pattern of PCMCO ($y = 0.3$). (b) the Co content dependence of the in-plane (blue squares), out-of-plane (red circles), and psudo-cubic (brown triangles) lattice parameters. Inset is an AFM image, the green line marks the lattice constant of the LSAT substrate.

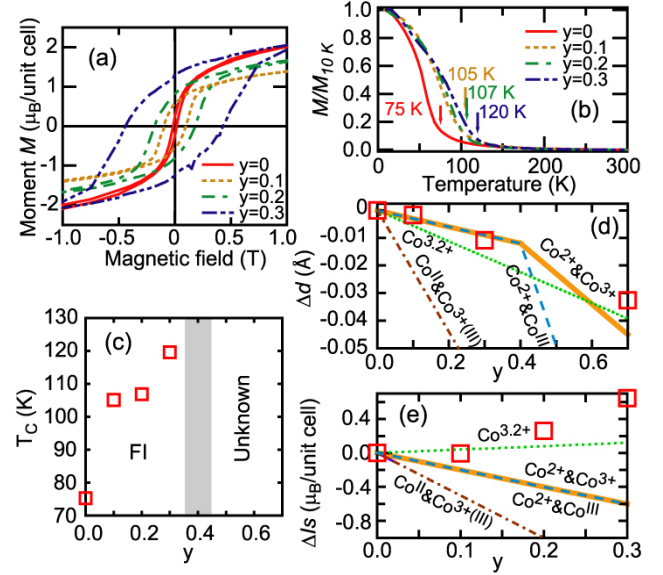

Fig. 2. Co content dependences of lattice constant and magnetic properties of epixtaxial PCMCO films. (a) M-H curves and (b) temperature dependences of magnetization (M-T curves) for samples with $0 \leq y \leq 0.3$. Arrows in (b) indicate the Curie temperature, defined here as the point where the tangent line through the inflexion point intercepts the temperature axis. (c) Curie temperature, (d) relative pseudo-cubic lattice constants and (e) relative saturation magnetization ($I_s$) referenced to non-substituted PCMO. In (d) and (e), the estimated lines for the substitution of $0.8Mn^{3+}+0.2Mn^{4+}$ for $Co^{2+}$ & $Co^{3+}$ (orange solid line), $Co^{2+}$ & $Co^{III}$ (blue dashed line), $Co^{II}$ & $Co^{3+(III)}$ (brown dot dash line), and $0.8Co^{3+}+0.2Co^{4+}$ (green dotted line, noted as average valency $Co^{3.2+}$) are shown with the experimental data (red squares).

According to the simple double exchange model, an increase of magnetic interactions between B-site ions should result in an increase of hopping conduction. The temperature dependence of resistivity was measured for the ferromagnetic samples ($0 \leq y \leq 0.3$). All samples show insulating properties that were robust up to an external magnetic field of 9 T, as shown in Fig. 3(a). This behavior is different from that of the metallic Co-substituted $Pr_{0.5}Ca_{0.5}MnO_3$ phase at low temperature[7]. Negative magnetoresistance, as is also seen in the conventional double exchange manganites, was observed as shown in Fig. 3(b). The magnetoresistance (MR) ratio measured above the Curie temperature increase with increasing Co content as shown in the inset of Fig. 3(b). MR below $T_c$ could not be measured due to the high resistivity of the PCMCO films.

The decrease of bond length and increase of the M-O-M bond angle by Co substitution [14, 15] and the increase of the hole-concentration in Mn-site induced by Co substitution[12] would result in increase of magnetic interaction between B-site ions according to the double exchange model. The increase of $T_C$, saturation magnetization, and MR ratio with increasing Co content in the $0 \leq y \leq 0.3$ ranges may be related to this effect. To confirm the increase of the hopping conduction in PCMCO, we carefully analyzed the temperature dependence of resistivity ($\rho$-$T$ curve) by fitting the data with variable range hopping (VRH) and Arrhenius models. The multiple correlation functions ($R^2$) of these fittings are plotted in the inset of Fig. 3(a). As Co content increases, the $R^2$ of the VRH model increase, which would enhance the double exchange interactions.

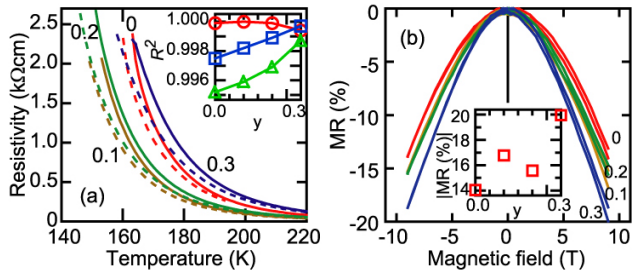

Fig. 3. Magnetotransport properties of epitaxial PCMCO films. (a) $\rho$-$T$ curves of the samples with $0 \leq y \leq 0.3$ under magnetic fields of 0 T (solid lines) and 9 T (dotted lines). Annotations indicate the Co content $y$. Inset shows the multiple correlation function $R^2$ of the fitting of the $\rho$-$T$ curves with an Arrhenius (red circles), 1-dimentional VRH (blue squares), and 3-dimentional VRH (green triangles) models. (b) Magnetic field dependences of the resistance. Inset shows the absolute value of MR under 9 T.

Besides double exchange interactions, another possible mechanism that contributes to the ferromagnetism of PCMCO is superexchange interactions. According to the Kanamori-Goodenough rule[16, 17], superexchange interactions for $Mn^{3+}$-O-$Mn^{3+}$, $Mn^{4+}$-O-$Mn^{3+}$, and $Mn^{4+}$-O-$Mn^{4+}$ favor antiferromagnetic coupling. By low-level, $0 \leq y \leq 0.3$, Co substitution, superexchange interactions for $Mn^{3+}$-O-$Co^{2+}$ and $Mn^{4+}$-O-$Co^{2+}$ can be formed. These include the ferromagnetic superexchange interactions for $Mn^{4+}$-O-$Co^{2+}$, which should contribute to the ferromagnetism. In highly substituted level ($y>0.4$), all the Mn ions would be $Mn^{4+}$ assuming that all the Co ions are divalent, resulting in vanishing of double exchange interaction between Mn ions. Alternatively, antiferromagnetic $Mn^{4+}$-O-$Mn^{4+}$, $Co^{2+}$-O-$Co^{2+}$, $Co^{2+}$-O-$Co^{3+}$, and $Co^{3+}$-O-$Co^{3+}$ interactions become dominant. These interactions compete with the $Mn^{4+}$-O-$Co^{2+(3+)}$ double exchange interactions and ferromagnetic $Mn^{4+}$-O-$Co^{2+}$ superexchange interactions in PCMCO, possibly forming unconventional magnetic phases with small magnetization as is observed in y=0.5, 0.7 (not shown).

In summary, we have studied the effect of Co substitution on the structural, magnetic, and transport properties of $Pr_{0.8}Ca_{0.2}MnO_3$ epitaxial thin films. Insulating properties were retained for all Co substitution levels. The detailed study on magnetic and transport properties suggest that both double exchange and superexchange interactions contribute to the ferromagnetic insulator phase in PCMCO, demonstrating the effectiveness of the B-site substitution to control the ferromagnetic properties of hole-doped perovskites by modulating both double exchange and superexchange interactions.

The authors acknowledge Dr. S. Nakao, Dr. Y. Hirose, and Prof. T. Hasegawa for MPMS and PPMS measurements. This work was supported by a Grant-in-Aid for Scientific Research S17101004, 20047003 and T. H. acknowledges financial support from JSPS.